\documentclass[aps,prb,showpacs,citeautoscript,twocolumn,superscriptaddress]{revtex4-1}

\usepackage{graphicx}
\usepackage{amsmath}
\usepackage{amssymb}
\usepackage{color}
\usepackage{subfigure}
\usepackage{braket}

\newcommand{\bea}{\begin{eqnarray*}}
\newcommand{\eea}{\end{eqnarray*}}
\newcommand{\bne}{\begin{equation*}}
\newcommand{\ede}{\end{equation*}}

\newcommand{\bnen}{\begin{equation}}
\newcommand{\eden}{\end{equation}}
\newcommand{\bean}{\begin{eqnarray}}
\newcommand{\eean}{\end{eqnarray}}
\newcommand{\bnsn}{\begin{subequations}}
\newcommand{\edsn}{\end{subequations}}

\newcommand{\bna}{\begin{array}}
\newcommand{\eda}{\end{array}}
\newcommand{\bnm}{\begin{enumerate}}
\newcommand{\edm}{\end{enumerate}}
\newcommand{\bni}{\begin{itemize}}
\newcommand{\edi}{\end{itemize}}

\newcommand{\threevector}[3]{\left(\bna{c} #1 \\ #2 \\ #3\eda\right)}

\renewcommand{\vec}[1]{\text{\boldmath{$ #1 $}}}

\begin{document}

\title{Control of valley dynamics in silicon quantum dots in the presence of an interface step}

\author{P\'eter Boross}
\affiliation{Institute of Physics, E\"otv\"os University, Budapest, Hungary}

\author{G\'abor Sz\'echenyi}
\affiliation{Institute of Physics, E\"otv\"os University, Budapest, Hungary}

\author{Dimitrie Culcer}
\affiliation{School of Physics, The University of New South Wales, Sydney 2052, Australia}

\author{Andr\'as P\'alyi}
\affiliation{Institute of Physics, E\"otv\"os University, Budapest, Hungary}
\affiliation{
Department of Physics and 
MTA-BME Condensed Matter Research Group,
Budapest University of Technology and Economics, Budapest, Hungary}

\date{\today}

\begin{abstract}
Recent experiments on silicon nanostructures have seen breakthroughs toward scalable, long-lived quantum information processing. The valley degree of freedom plays a fundamental role in these devices, and the two 
lowest-energy electronic states of a silicon quantum dot can form a valley qubit. 
In this work, we show that a single-atom high step at the silicon/barrier
interface induces a strong interaction of the qubit and 
in-plane electric fields, and analyze the consequences 
of this enhanced interaction on the dynamics of the qubit. 
The charge densities of the qubit states are deformed differently by the interface step, allowing non-demolition 
qubit readout via valley-to-charge conversion.
A gate-induced in-plane electric field together with
the interface step enables fast control of the valley qubit 
via electrically driven valley resonance. 
We calculate single- and two-qubit gate times, as well as relaxation and dephasing times, and present predictions for the parameter range where the gate times can be much shorter than the relaxation time and dephasing is reduced.
\end{abstract}

\maketitle

\section{Introduction}

Localized spins in silicon quantum dot (QD) and donor systems are actively investigated as platforms for quantum computing \cite{Kane_Nature98, Zwanenburg_SiQmEl_RMP13}. The chief reason for this is their long spin coherence times \cite{Feher_PR59, FeherGere_PR59, Roth_PR60, Hasegawa_PR60, Eriksson_SiQbt_Loading_10, Raith_SiQD_1e_SpinRelax_PRB11, Tahan2014, Berme, Kha_APL15} due to weak spin-orbit coupling, the existence of nuclear-spin free isotopes allowing isotopic purification \cite{Itoh_MRS}, and the absence of piezoelectric electron-phonon coupling \cite{Yu_Cardona}. Recent years have witnessed enormous experimental strides towards making silicon quantum computing scalable and long-lived \cite{Morello_1glShot_Nature10, Tracy_MOSFET_TunableDQD_APL10, Pla-electronspin, Pla-nuclearspin, Veldhorst_addressable}, with long spin coherence times observed for single electrons, \cite{Muhonen_Store_14} as well as demonstrations of electrical spin control \cite{Kawakami2014} and entanglement \cite{Veldhorst_twoqubit, Weber_NN14}. 

The valley degree of freedom has emerged as an important ingredient of silicon quantum bits (qubits). It increases the size of the qubit Hilbert space and introduces fundamental complications in particular in entanglement, such as exchange oscillations in donors and suppression in QDs \cite{Cullis,Koiller_PRL01, Friesen_PRB04, Wellard_PRB05, Friesen_PRB07, Culcer_PRB10, Friesen_PRB10}. For example, valley interference effects have recently been experimentally observed in donors \cite{Salfi_VlyInt_NM14}. In addition, intervalley spin-orbit coupling terms can induce simultaneous spin-valley dynamics, affecting spin relaxation as well as the $g$-factor of QDs \cite{Kawakami2014,Nestoklon_PRB06, Dery_Si_SO_10, Peihao_SpinValley_14, Veldhorst_IntVlySO_PRB15}. Interestingly, the valley splitting can be measured and the valley degree of freedom can be controlled by a means of a gate-induced out-of-plane electric field \cite{Saraiva_EMA_PRB11, Lim_SiQD_SpinFill_NT11, Wu_Culcer_PRB2012, Culcer_ValleyQubit_PRL12, Yang_SpinVal_NC13,DohunKim_hybridqubit, Hao_SiDQD_SpinVlly_NC14}. This realization has led to the proposal of a two-electron qubit encoded in the valley degree of freedom itself \cite{Culcer_ValleyQubit_PRL12}, which is expected to have good coherence properties \cite{Tahan2014}.
Quantum control and coherence properties 
of valley qubits and combined spin-valley 
qubits are also being explored actively in a range of other 
materials, including graphene\cite{Recher}, 
carbon nanotubes\cite{Palyi_PRL11,Laird-nn,Szechenyi-maximalrabi,Laird-review}, 
and transition metal dicalcogenides\cite{Kormanyos,YueWu}. 

\begin{figure}
\begin{center}
\includegraphics[width=1\columnwidth]{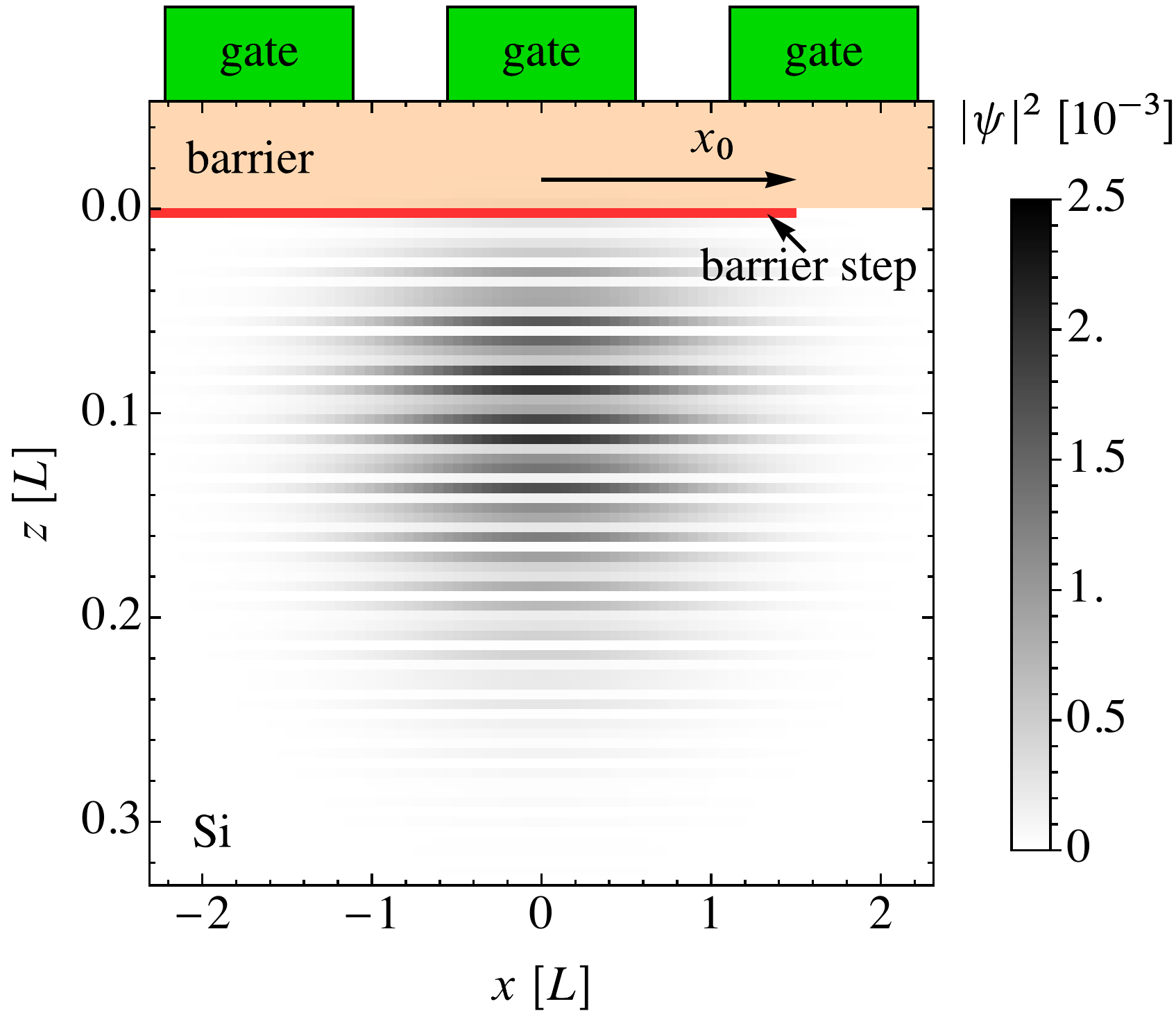}
\end{center}
\caption{(Color online) 
Particle density of a single electron in a silicon 
quantum dot
in the presence of a single-atom high step
at the silicon/barrier interface. 
Lateral confinement in the $xy$ plane is parabolic
and centered at the origin.
Light orange (light gray) region, $z<0$: barrier material. 
Red (dark gray) stripe: a single-atom high step consisting of the
barrier material, assumed to be translationally invariant
along $y$. 
The relative position of the step edge and the center of the 
lateral confinement potential is denoted by $x_0$.
}
\label{fig:setup}
\end{figure}

In this work, we theoretically study the dynamics 
of a single-electron valley qubit in a silicon QD. 
The valley qubit is formed by the two lowest-energy 
electronic states of the QD. 
We demonstrate that a single-atom high step at the silicon/barrier
interface (see Fig.~\ref{fig:setup}), 
a defect ubiquitous in silicon nanostructures, 
can induce a strong interaction of the qubit and 
in-plane electric fields.
We show that the charge densities of the two qubit states are deformed differently by the interface step, as the relative position of the
QD and the edge of the interface step is tuned by 
a gate-induced in-plane electric field.
This provides an opportunity for non-demolition qubit readout
via valley-to-charge conversion. 
Furthermore, we demonstrate that, in the vicinity of the step, the physics of the valley qubit is analogous to that of a charge qubit in a double quantum dot.

Our main goal then is to discuss and quantify the coherent-control
opportunities and the decoherence mechanisms arising from 
the enhanced interaction between the qubit and the in-plane electric
fields.
We determine the relaxation and dephasing matrix elements 
characterizing this interaction.
We discuss the role of the relaxation matrix element in enabling fast single-qubit control via electrically driven valley resonance as well as entanglement via an $\sqrt{i\rm SWAP}$ two-qubit gate. 
Concomitantly, we study qubit relaxation via spontaneous phonon emission and show that, although valley relaxation times can range over several orders of magnitude, in certain parameter regimes the single-qubit gate times can be much shorter than the relaxation time, allowing approximately $10^3$ operations in one relaxation time. Finally, we discuss qubit dephasing rates due to background charge fluctuations and identify an operational window in which dephasing is reduced.

We model the system based on the hybrid approach of Ref. \onlinecite{Gamble2013}, using the effective mass approximation to describe the dynamics in the plane of the interface and a tight-binding approximation for the dynamics perpendicular to the interface. 
Our setup does not include a magnetic field and we do not consider explicitly the spin degree of freedom.

The outline of this paper is as follows. In Sec.\ \ref{sec:setupandmodel} we introduce the physical setup considered in this work and the model Hamiltonian used to study it, while in Sec.\ \ref{sec:valleytocharge} we discuss in detail valley-to-charge conversion. Sec.\ \ref{sec:vqd} is devoted to the dynamics of the valley qubit, comprising coherent control due to an external electric field as well as relaxation due to phonons and dephasing due to charge noise. We summarize our findings in Sec.\ \ref{sec:sum}.

\section{Setup and model}
\label{sec:setupandmodel}

We consider a single conduction-band 
electron in a gate-defined QD at a 
silicon/barrier interface.
The setup, along with the spatial dependence of the 
ground-state
particle density of the electron, is shown in Fig.~\ref{fig:setup}.
The colored regions represent
the barrier material (e.g., SiGe or SiO$_2$), 
and the region below represents
silicon.
The  $z$ axis is aligned with the [001] crystallographic 
direction of silicon. 
The in-plane confinement potential for the electron is created by top gates.
The gate-induced electric field pushes
 the electron against the barrier, 
and also defines the lateral confinement parallel to the
$xy$ plane. 
The key element of the setup is a single-atom high barrier 
step at the silicon/barrier interface,
depicted as the red (dark gray) stripe. 
The step consists of the atoms of the barrier material,
is assumed to be translationally invariant along y, 
and the relative position of the step edge and the 
center of the lateral confinement potential is denoted by $x_0$. 

Note that the setup considered here, incorporating a half-infinite
barrier step at the silicon/barrier interface, 
is similar to the one considered 
in Ref.~\onlinecite{Gamble2013}.
(The model used in Ref.~\onlinecite{Gamble2013}
is also adopted here, see below.)
Therein, the authors describe a single-atom
high barrier step that has a rectangular shape in the $xy$ plane, 
with a fixed location 
with respect to the lateral confinement potential, 
and compute and discuss how the energy levels and
the dephasing matrix elements (see below for definitions)
depend on the
gate-induced $z$-directional 
electric field that pushes the electron
toward the upper barrier.
Inspired by that study, here we address the
following distinct questions:
(i) How do the wave functions behave as the relative position
of the step and the QD is varied, for example, under the action of an in-plane electric field?
(ii) What is the reason for the observed behavior? 
(iii) What are the qualitative and quantitative 
consequences of the observed behavior 
in the context of coherent control and information loss
of the valley qubit?

We describe the electron using the hybrid model introduced in 
Ref.~\onlinecite{Gamble2013},
which combines the envelope-function approximation 
to treat the wave function in the $xy$ plane, with 
a one-dimensional 
tight-binding model\cite{Friesen_PRB04} (\emph{chain}) 
along the  $z$ axis.
That distinction between the $xy$ plane and the  $z$ direction is made 
to account for the fact 
that the electronic wave function in a silicon quantum dot
is a packet of Bloch waves that reside
in the $z$ and $\bar{z}$ valleys of silicon's conduction band. 
As a consequence, in this hybrid model, the 
electronic wave function 
$\psi(x,y,j)$
has two continuous spatial variables,
the $x$ and $y$ coordinates,  
and one 
integer spatial variable, the site index $j$ along the chain.
Here, the integer $j$ 
is associated to the position $z_j = (j-1/2)a$ of the $j$th site
of the chain,
and $a$ is the lattice constant of the latter.
We use the normalization relation
$\int dx \int dy \sum_j |\psi(x,y,j)|^2 = L^2$, where $L$ is the lateral confinement length defined below.
Of course, the spatial structure of the Hamiltonian is
analogous to that of the wave function.

\begin{figure}
\begin{center}
\includegraphics[width=1\columnwidth]{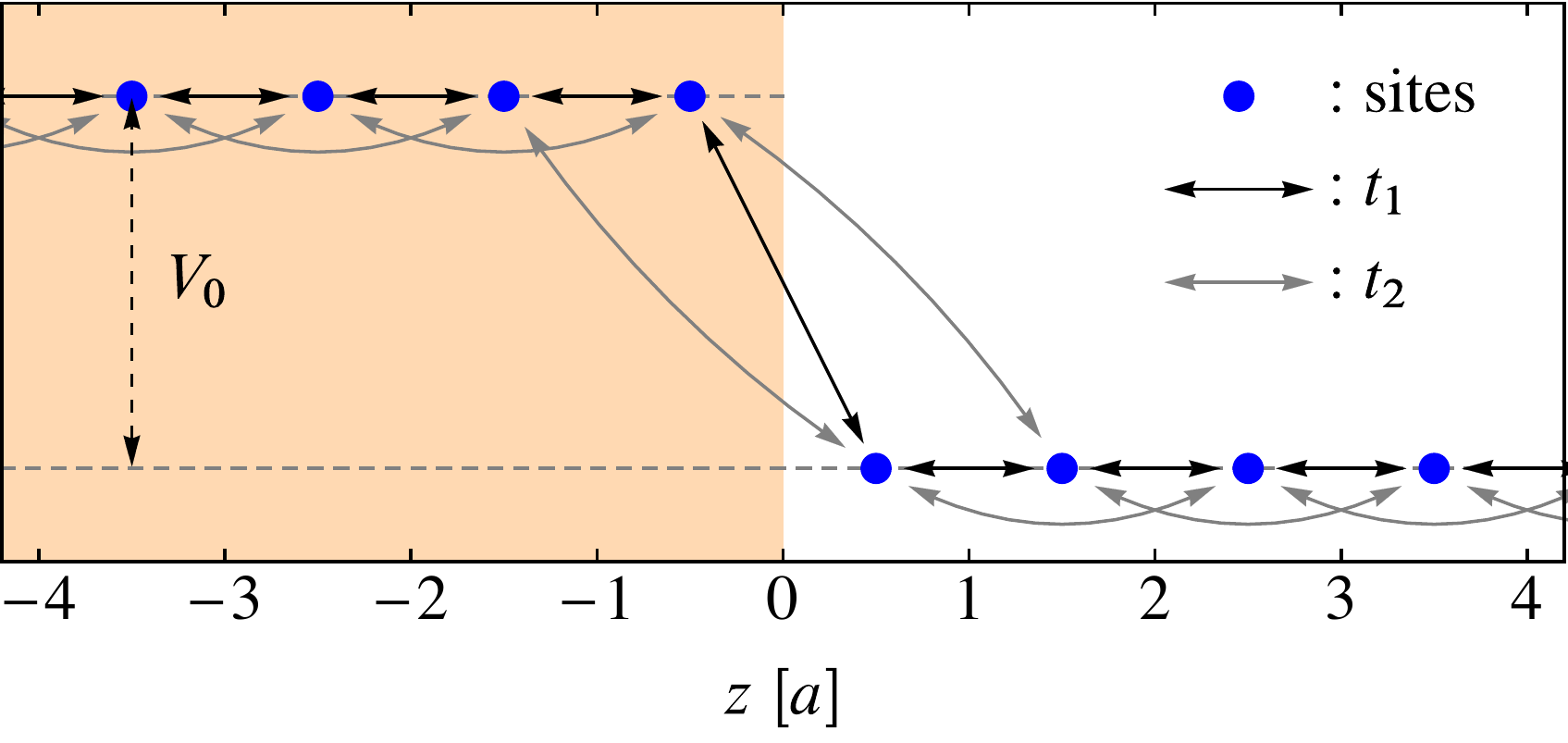}
\end{center}
\caption{(Color online)
One-dimensional tight-binding model near the interface along $z$. 
Nearest-neighbor ($t_1$) and next-nearest-neighbor ($t_2$)
hopping amplitudes are shown as black (gray) arrows. 
The shaded region is the barrier material. 
The vertical axis represents the on-site energy; 
the conduction-band offset $V_0$ of the barrier material  is shown 
but the interface electric field is not.}
\label{fig:chain}
\end{figure}

In the chain along $z$, depicted in Fig.~\ref{fig:chain}, the neighboring
sites represent neighboring atomic layers of silicon,
therefore $a = a_0/4$ is chosen, with $a_0 = 0.543\ \text{nm}$ being
the lattice constant of silicon. 
Note also that we neglect the spin degree of freedom from now on. 

The electron in the QD is confined by 
the gate-induced electric fields and by 
the barrier material. 
These effects are taken into account as electrostatic
potentials:
\begin{subequations}
\label{eq:potentials}
\bean
V &=& 
V_i + V_{xy} + V_b,\\
V_i(j)  &=& e E_z z_j, \\
V_{xy}(x,y) &=& \frac 1 2 m_{xy} \omega_0^2(x^2+y^2), \\
V_b(x,y,j) &=& V_0 \chi_b(x,y,j).
\eean
\end{subequations}
Here, $V_i$ represents the 
interface electric field $E_z$
pushing the electron against the barrier,
$V_{xy}$ represents the gate-induced lateral confinement potential,
and $V_b$ represents the conduction-band offset of the 
barrier material ($V_0$). 
The function $\chi_b$ specifies the spatial range of the 
barrier material: 
$\chi_b(x,y,j < 1) = 1$, 
$\chi_b(x,y,j > 1) = 0$, 
and $\chi_b(x,y,j=1) = \Theta(x_0 - x)$,
where $\Theta$ is the Heaviside function.
Furthermore, $m_{xy} = 0.19\, m_0$ is the transverse effective mass
of the silicon conduction band.
We also introduce the lateral confinement length $L = \sqrt{\hbar/(m_{xy} \omega_0)}$.

The complete Hamiltonian $H= K+ V$ also incorporates
the kinetic energy term $K$:
\bnen
\label{unpert}
K = \frac{p_x^2+p_y^2}{2m_{xy}}+K_\text{chain}.
\eden
Here, the kinetic energy
associated to electron hopping between atomic layers along 
the  $z$ direction: 
\bnen
\left(K_\text{chain}\right)_{i,j} = t_1(\delta_{i,j+1}+\delta_{i,j-1})+t_2(\delta_{i,j+2}+\delta_{i,j-2}),
\eden
where $t_1=683\ \text{meV}$  ($t_2 = 612\ \text{meV}$) is the nearest-neighbor (next-nearest-neighbor)
hopping amplitude. 
These values are set\cite{Friesen_PRB04} so that
the corresponding one-dimensional
bulk dispersion relation reproduces the longitudinal
effective mass $m_z = 0.916 \, m_\text{e}$ 
of the conduction-band bottom as well as
the momentum  $z$ component
$k_0 = 0.82  \, (2\pi / a_0)$
corresponding to the 
 $z$ and $\bar{z}$ valleys of silicon's conduction band.

In what follows, we focus on the properties of the lowest two
energy eigenstates of the QD, $\ket{\text{g}}$ and $\ket{\text{e}}$,
which are computed numerically by diagonalizing
the complete Hamiltonian $H$.
We set $V_0 = 150\ \text{meV}$, representing
SiGe as the barrier material.
The results presented here were obtained using an interface
electric field of 
$E_z = 3\ \text{MV/m}$, and 
a lateral 
confinement energy $\hbar \omega_0 = 0.5\ \text{meV}$, corresponding to 
a lateral confinement length $L \approx  28.3\ \text{nm}$.
Further details of the model and the numerical implementation
are in Appendix \ref{app:model}.

\begin{figure*}
\begin{center}
\includegraphics[width=\textwidth]{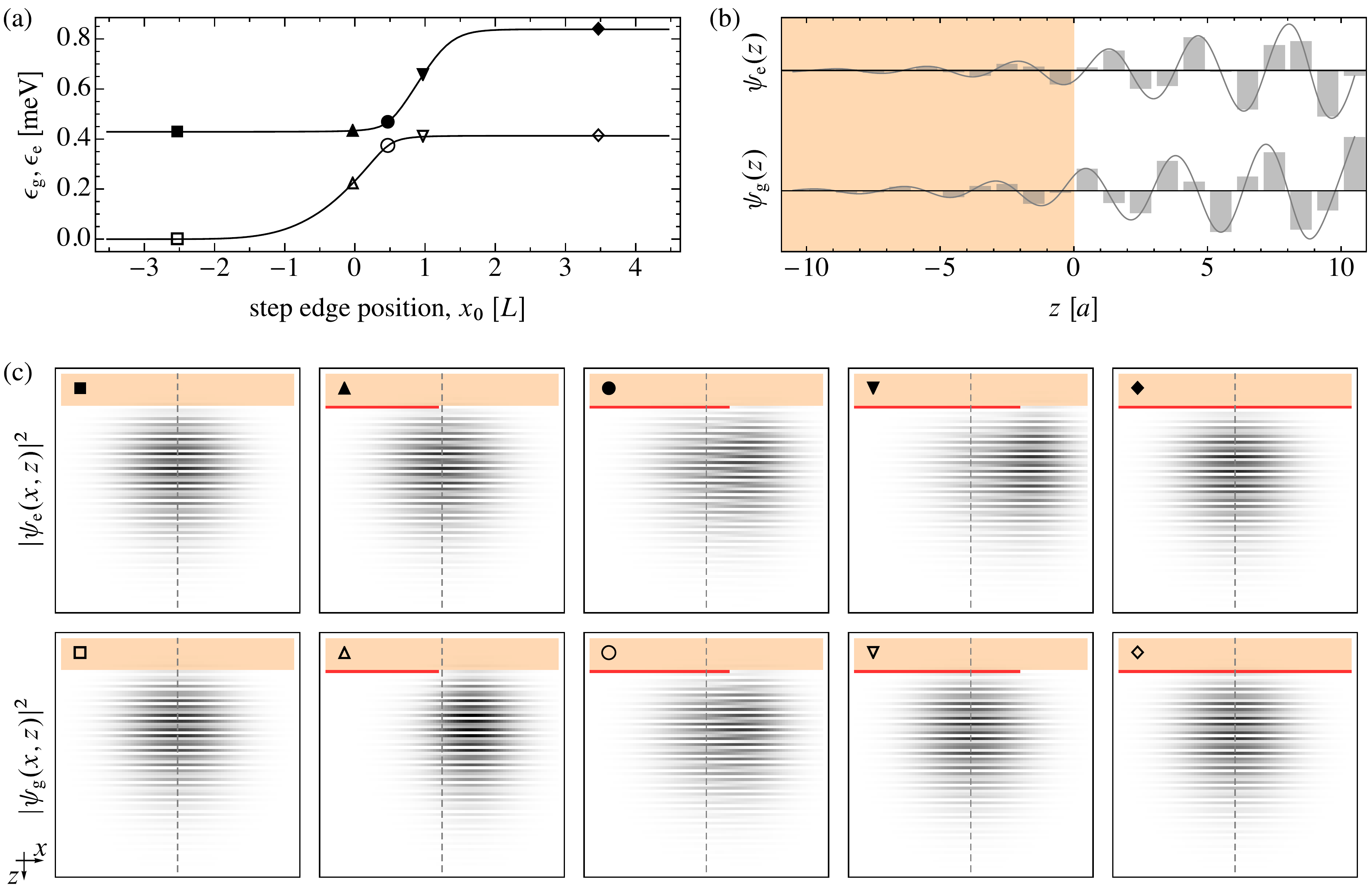}
\end{center}
\caption{(Color online) 
Energies and wave functions
of a single electron in a silicon quantum dot.
(a) Energy eigenvalues $\epsilon_\text{g}$ and $\epsilon_\text{e}$ 
of the ground state and the first excited state, as functions
of the step edge position $x_0$.
(b) Bar chart:  $z$ dependence of the 
ground-state $\ket{\text{g}}$ and excited-state $\ket{\text{e}}$ 
wave functions 
in the absence of the step. 
The solid line is a guide to the eye, 
smoothly connecting the data points to highlight
the wave-function oscillations with wave number $k_0$. 
(c) Particle densities of the ground state $\ket{\text{g}}$ (bottom row)
and excited state $\ket{\text{e}}$ (top row)
at selected values of the step edge position $x_0$. 
Reference frame, length scales and grayscale 
are defined in Fig.~\ref{fig:setup}.
Dashed horizontal lines denote $x=0$, 
the center of the lateral confinement potential along $x$.
}
\label{fig:wfn}
\end{figure*}

\section{Valley-to-charge conversion}
\label{sec:valleytocharge}

We  consider the two-level system formed by the
two lowest-energy eigenstates, $\ket{\text{g}}$ and $\ket{\text{e}}$, of 
the Hamiltonian $H$.
We  refer to this system as the valley qubit,
and to the two energy eigenstates as the valley-qubit basis states.
If the QD is located on a flat silicon/barrier interface, 
then the gross spatial features of the
charge densities of the two valley-qubit basis states are
very similar (see, e.g., Fig.~\ref{fig:wfn}c, leftmost column), 
indistinguishable for a usual charge sensor
that lacks atomic spatial resolution. 
In this section, we argue that a single-atom high interface step 
can be utilized to bring the valley qubit to a state where
it resembles a conventional charge qubit in a double quantum dot. We refer to this phenomenon as
\emph{valley-to-charge conversion}.
Therefore, in principle, such a setup allows 
for a projective non-demolition readout of the 
valley qubit via charge sensing. 
Furthermore, in the next section we quantify how
this valley-to-charge conversion strengthens the
interaction of the valley qubit with electric fields, 
and, in turn, how it enhances the effectivity of 
coherent-control operations as well as decoherence
mechanisms. 

Let us start by introducing the key parameter $x_0$, 
which we refer to as the step edge position. 
In the considered setup, see Fig.~\ref{fig:setup},
we identify the origin
of the $x$ axis with
the center of the lateral confinement potential.
The step edge position $x_0$ is 
defined as the distance between the center of the
lateral confinement potential 
and the step edge.
We envision the possibility that $x_0$ is 
in situ tuneable: 
a sufficiently sophisticated 
top-gate electrode structure could be utilized to control $x_0$
by moving the lateral QD confinement potential and hence the electron
itself along the $x$ axis.

The opportunity for valley-to-charge conversion is
suggested by the  $z$ dependence of the wave 
functions of $\ket{\text{g}}$ and
$\ket{\text{e}}$, in the absence of the step\cite{Friesen_PRB07,Tahan2014}.
In this case, the wave function is a product of $x$-, $y$- and 
$z$-dependent factors; 
the dependence of $\ket{\text{g}}$ and $\ket{\text{e}}$ on  $z$ is 
shown in Fig.~\ref{fig:wfn}b
(cf.~Fig.~8 of Ref.~\onlinecite{Tahan2014}).
The key observations are as follows.
(i) The 
ground state $\ket{\text{g}}$ has a nearly vanishing
wave function at the
last atomic layer of the barrier material ($z_j=-a/2$),
and it has a peak
in the first silicon layer ($z_j=a/2$).
(ii) The wave function of the excited state $\ket{\text{e}}$
is peaked at the last barrier layer but is close to zero 
at the first silicon layer. 
A simple interpretation of these two observations is 
given in Appendix \ref{sec:interpretation}.

As a consequence of (i),
we expect that if the electron occupies
$\ket{\text{g}}$, and we try to move it above the step, 
then it will resist to
move together with the lateral  confinement potential,  
as it has an 
appreciable probability of being in the atomic layer of the step.
However, as a consequence of (ii), if the electron occupies 
$\ket{\text{e}}$, then it can follow the lateral confinement potential
as its probability of occupying the layer of the atomic step
almost vanishes.

To confirm these expectations, we computed 
the wave functions of the 
energy eigenstates $\ket{\text{g}}$ and $\ket{\text{e}}$.
The corresponding particle densities in the $y=0$ plane
are shown in Fig.~\ref{fig:wfn}c, 
for five selected values of the step 
edge position $x_0$, see Fig.~\ref{fig:wfn}a, confirming our expectations
about valley-to-charge conversion.
The subplots $\square$ and $\blacksquare$ 
of Fig.~\ref{fig:wfn}c
show the ground-state and excited-state 
particle densities when the electron 
is confined far from the step, $x_0 \approx -1.8 L$. 
As an attempt is made to move the electron on the step by the gates,
that is, the lateral confinement potential is moved 
such that $x_0  \approx 0$ (subplot $\vartriangle$ and $\blacktriangle$
in Fig.~\ref{fig:wfn}c), 
the ground-state electron
is stuck on the right side of the step ($\vartriangle$),
whereas the excited-state electron moves onto
the step ($\blacktriangle$), in accordance with the argument of the 
previous paragraph.
Therefore, by moving the lateral confinement potential 
to this position, the valley-to-charge conversion 
has been completed, and a charge-sensing measurement
at this position could provide a projective non-demolition 
readout of the valley qubit.  

The dependence of the
energy eigenvalues $\epsilon_\text{g}$, $\epsilon_\text{e}$ 
of the valley-qubit basis states
on the step edge position $x_0$ is shown in Fig.~\ref{fig:wfn}a.
The two energy eigenvalues exhibit a familiar anticrossing pattern,
located around $x_0 \approx 0.5 L$:
the first energy eigenvalue, 
corresponding to the ground state for $x_0 < 0$, 
moves upwards as $x_0$ increases, 
and anticrosses with the apparently flat second energy eigenvalue.
Using the above observations (i) and (ii), a straightforward 
interpretation of this pattern can be given.
As the ground-state electron is pushed against the step, 
it does feel the presence of the step [see (i)], and 
therefore its confinement along $x$ gets tighter 
and its wave function along $x$ gets squeezed
(Fig.~\ref{fig:wfn}c, $\vartriangle$); thereby 
its energy increases. 
As the excited-state electron is pushed against the step,
it hardly feels its presence [see (ii)], therefore its charge
center follows the center of the lateral confinement potential, 
the shape of its wave function along $x$ remains intact
to a good approximation (Fig.~\ref{fig:wfn}c, $\blacktriangle$), 
and its energy remains 
essentially unchanged. 
When these two energy eigenvalues meet at $x_0 \approx 0.5 L$,
an anticrossing opens because the potential representing
the step provides a nonzero coupling matrix element between 
the two states $\blacktriangle$ and
$\vartriangle$.

In the vicinity of this anticrossing point at $x_0 \approx 0.5 L$, 
the behaviour of the valley qubit, as a function 
of the step edge position $x_0$,
strongly resembles the behavior
of a charge qubit in a double quantum
dot (DQD), as a function of its detuning parameter $\varepsilon$. 
Here, detuning $\varepsilon$ and tunnel coupling $t$ 
are the two parameters in the charge-qubit Hamiltonian
\bean
\label{eq:chargequbit}
H_\text{cq} = - \frac 1 2 \varepsilon \sigma_3 + \frac 1 2 t \sigma_1,
\eean 
with $\sigma_j$ representing the Pauli matrices
in the (left dot, right dot) two-dimensional Hilbert space.
The features supporting the 
analogy between the valley qubit and the charge qubit
are as follows. 
(i) At $x_0 \approx 0$, the two valley qubit basis states
($\vartriangle$ and $\blacktriangle$ in Fig.~\ref{fig:wfn}c),
are well
localized and separated
from each other.
This corresponds to the charge qubit  at
$\varepsilon < - |t|$.
(ii) At the anticrossing point $x_0 \approx 0.5$,
the valley-qubit energy splitting has a minimum,
similarly to the charge qubit
energy splitting $\sqrt{\varepsilon^2 + t^2}$ at $\varepsilon = 0$. 
The particle densities of the two valley qubit basis states
($\circ$ and $\bullet$ in Fig.~\ref{fig:wfn}c), 
are rather delocalized and hardly
indistinguishable;
essentially, they are bonding and antibonding combinations of
the ones in Fig.~\ref{fig:wfn}c $\vartriangle$ and $\blacktriangle$, 
analogous to the eigenstates $(1,1)/{\sqrt{2}}$
and $(1,-1)/{\sqrt{2}}$ of the charge qubit Hamiltonian 
$H_\text{cq}$ at zero detuning $\varepsilon = 0$. 
(iii) On the other side of the anticrossing, around
$x_0 \approx L$, 
the two valley qubit basis states
($\triangledown$ and $\blacktriangledown$ in Fig.~\ref{fig:wfn}c),
swap their location with respect to 
(i), and are again well localized and separated
from each other. 
This corresponds to the charge qubit  at
$\varepsilon > |t|$.

\section{Valley-qubit dynamics}\label{sec:vqd}

\newcommand{\xrme}{\braket{\text{e}|x|\text{g}}}
\newcommand{\zrme}{\braket{\text{e}|z|\text{g}}}
\newcommand{\xdme}{\braket{\text{e} | x |\text{e}} - \braket{\text{g}| x |\text{g}}}
\newcommand{\zdme}{\braket{\text{e} | z |\text{e}} - \braket{\text{g}| z |\text{g}}}

Our central goal is to describe the influence of the
step on  valley-qubit dynamics,
including coherent qubit control via external electric fields,
as well as information loss processes.
Key quantities enabling the 
quantitative characterization of those are the
relaxation and dephasing matrix elements (see below for definitions).
These matrix elements indicate
the strength of the interaction of the qubit with
electric fields. 
As a generic conclusion, we will show that this interaction
is strongly enhanced by the presence of the step. 

In the next subsection, we analyze the
behavior of the relaxation and dephasing matrix elements
as the 
function of the step edge position $x_0$, 
and show that around the
valley-qubit energy anticrossing point, 
their behavior is analogous to the
relaxation and dephasing matrix elements 
of a DQD charge qubit around zero detuning $(\varepsilon =0)$.
Then, partly relying on the relaxation and dephasing matrix elements,
we complete our goal by analyzing the way the presence of the
step speeds up coherent qubit operations and
information loss processes.

\subsection{Interaction with an electric field: 
relaxation and dephasing matrix elements}
\label{sec:interaction}

\begin{figure*}
\begin{center}
\includegraphics[width=\textwidth]{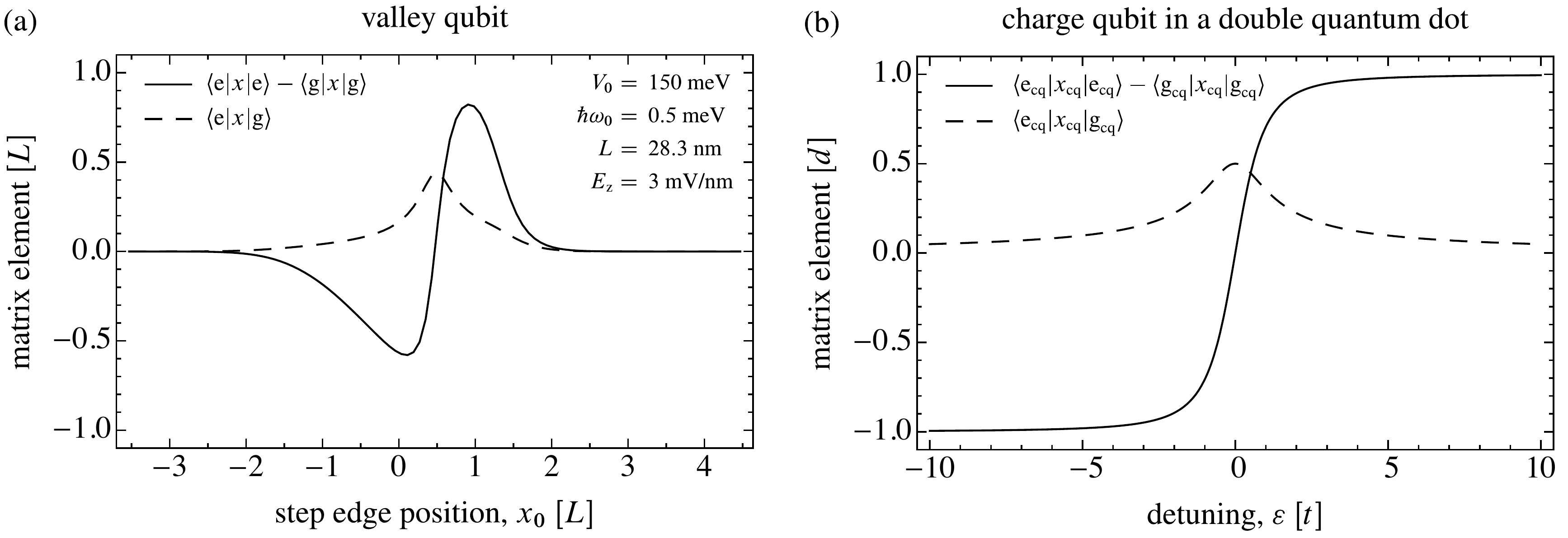}
\end{center}
\caption{(Color online) 
Relaxation and dephasing matrix elements:
analogy between the valley qubit and the charge qubit.
(a) Dependence of the
$x$-directional relaxation (dashed) and dephasing (solid)
matrix elements on the step edge position for the valley qubit. 
(b) Dependence of the relaxation (dashed) and 
dephasing (solid) matrix elements on the on-site energy
detuning $\varepsilon$ of a single-electron charge qubit
in a double quantum dot . 
}
\label{fig:matrixelements}
\end{figure*}

We consider the valley qubit in the presence of a step; 
the system is described by the Hamiltonian $H$ introduced above. 
We assume that there is an additional, weak, 
potentially time-dependent,
homogeneous electric field $E_x(t)$ along $x$, induced intentionally by
applied voltages on the gates or unintentionally by noise;
its effect is described via the electric-field Hamiltonian 
$H_E(t) = e E_x(t) x$.

A simple way to 
describe the effect of this electric field on the qubit dynamics
is via the effective Hamiltonian
$H_\text{vq} = \tau_0 [H +H_E(t)] \tau_0 $ of the valley qubit,
that is, the projection
of the complete Hamiltonian onto the two-dimensional 
valley-qubit subspace of $H$, using 
$\tau_0  = \ket{\text{g}}\bra{\text{g}}+\ket{\text{e}}\bra{\text{e}}$.
The effective Hamiltonian of the valley qubit reads
\bean
\label{eq:electricfield}
H_\text{vq} 
&=& 
\frac 1 2 \hbar \omega_\text{vq} \tau_3
\\
\nonumber
&+&
e E_x(t) \left[
	\xrme \tau_1 +
	\frac{\xdme}{2} \tau_3
\right].
\eean
Here, 
$\omega_\text{vq} = (\epsilon_\text{e} - \epsilon_\text{g})/\hbar$ 
is the Larmor frequency of the valley qubit, and
$\tau_j$ ($j=1,2,3$) are the Pauli matrices acting in the valley-qubit
subspace, e.g., $\tau_3 = \ket{\text{e}} \bra{\text{e}} - \ket{\text{g}}\bra{\text{g}}$. 
Equation \eqref{eq:electricfield} testifies that the interaction
between the valley qubit and the electric field is
characterized by the quantities
$\xrme$ and $\xdme$; we refer to those as the
$x$-directional relaxation matrix element 
and dephasing matrix element, 
respectively.

Note that we choose the energy eigenstates of $H$
as real-valued functions.
This ensures that not only the dephasing matrix element 
but also the relaxation matrix element is real valued.
The sign of the relaxation matrix element is still ambiguous,  
but this has no physical relevance.

The numerically computed $x$-directional 
relaxation and 
dephasing matrix elements,
as functions of the step edge position $x_0$, 
are shown in Fig.~\ref{fig:matrixelements}a.
The relaxation matrix element (dashed line in 
Fig.~\ref{fig:matrixelements}a) is small
when the QD and the step does not overlap, 
and shows 
a peak at the anticrossing point $x_0 \approx 0.5 L$, with a
height of $\approx L/2$ and a full width at half 
maximum of $\approx L$.
The dephasing matrix element (solid line in 
Fig.~\ref{fig:matrixelements}a) is also small 
when the QD and the step are far away from each other, 
has a minimum and a maximum on the two sides of the 
anticrossing point, and vanishes at the anticrossing point.
Note that since the dephasing matrix element measures
the distance of the charge centers of $\ket{\text{g}}$, its
qualitative behavior is seen already from the wave functions
in Fig.~\ref{fig:wfn}c.

Around the anticrossing point $x_0 \approx 0.5 L$, 
where the relaxation and
dephasing matrix elements are sizeable, their 
behavior is similar to those 
of a DQD 
charge qubit around  zero detuning $E_z = 0$.
In our minimal model of the charge qubit,
see Eq.~\eqref{eq:chargequbit},
the position operator
is represented as $x_\text{cq} = - d \sigma_3/2$, where
$d$ is the spatial separation between the centers of the 
two QDs that are placed along the $x$ axis. 
Therefore, the relaxation and dephasing matrix elements of
the charge qubit read
\begin{subequations}
\bean
\braket{\text{e}_\text{cq} | x_\text{cq} |\text{g}_\text{cq} } &=&
\frac{t}{2\sqrt{t^2+\varepsilon^2}} \, d,
\\
\braket{\text{e}_\text{cq} | x_\text{cq} |\text{e}_\text{cq}} 
-
\braket{\text{g}_\text{cq} | x_\text{cq} |\text{g}_\text{cq}} 
&=& \frac{\varepsilon}{\sqrt{t^2 + \varepsilon^2}}\,  d,
\eean
\end{subequations}
Here, $\ket{\text{g}_\text{cq}}$ and $\ket{\text{e}_\text{cq}}$ are the
ground and excited states of the charge-qubit Hamiltonian
$H_\text{cq}$, respectively.
The relaxation and dephasing 
matrix elements of the charge qubit are shown in 
Fig.~\ref{fig:matrixelements}b.
Comparing the trends of the matrix elements of the two 
qubits, the only notable qualitative difference is
that the dephasing matrix element of the valley qubit approaches
zero away from the anticrossing point.
This is intuitively obvious: the dephasing matrix element characterizes the
spatial separation of the charge centers of the two states, 
which indeed approaches 
zero for the valley qubit if the QD is placed
at a large distance from the step.

The $y$- and $z$-directional relaxation 
and dephasing matrix elements
are defined analogously to the $x$-directional ones.
Our numerical results confirm the observation\cite{Gamble2013} that 
the $z$-directional matrix elements are of the order of the
lattice constant $a$.
Recall that the typical scale of the $x$-directional matrix element
is the lateral dot size $L \gg a$;
this implies that the role of the $z$-directional matrix elements
in the step-induced valley-qubit dynamics
is marginal.
Therefore, even though they are taken into account in the calculations,
they are disregarded in the upcoming discussion.
Finally, 
the $y$ dependence of the wave functions of $\ket{\text{g}}$ and $\ket{\text{e}}$
separates from the $x$ and  $z$ dependencies, 
and takes the form of the 
Gaussian ground state of the parabolic confinement
potential along $y$, hence the $y$-directional relaxation and dephasing
matrix elements vanish. 

\subsection{Coherent control of a single valley qubit via
electrically driven valley resonance}

One important conclusion drawn from the previous subsection
is that the interaction between the valley qubit and electric fields
gets strongly enhanced when the QD is in the vicinity of the
interface step. 
Here we argue that this enhanced interaction can be utilized
to coherently control the valley qubit with an ac electric field
in a resonant fashion (\emph{electrically driven valley resonance}). 
Controlling the valley qubit with an ac electric field is 
similar 
to the electrically driven spin (valley) resonance mechanism
in semiconductor \cite{Nowack_Science07,Golovach_EDSR_PRB06} 
(carbon nanotube \cite{Palyi_PRL11}) 
QDs, and can be triggered by an ac voltage component applied on
one of the confinement gates.

\begin{figure}
\begin{center}
\includegraphics[width=1\columnwidth]{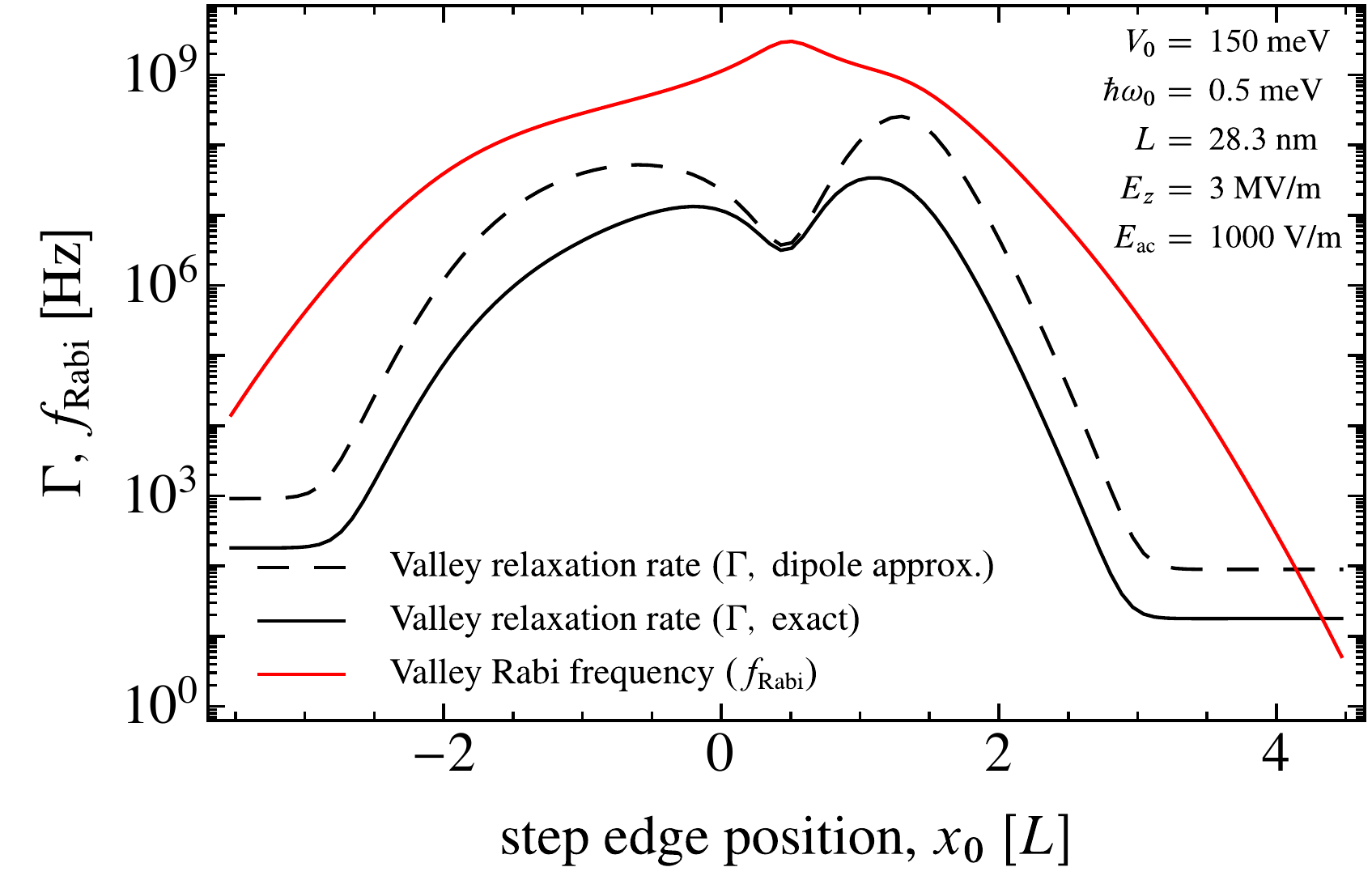}
\end{center}
\caption{(Color online) 
Coherent control and phonon-mediated relaxation
of the valley qubit.
Red (gray) solid: valley Rabi frequency as a function of the
step edge position, for a driving electric field $E_\text{ac} = 1000\ \text{V/m}$. 
Black dashed/solid: zero-temperature relaxation rate
of the valley qubit evaluated with/without the dipole approximation.
}
\label{fig:rabirelaxation}
\end{figure}

The fact that an $x$-directional ac electric field can drive coherent
Rabi oscillations of the valley qubit is a simple consequence of the 
effective Hamiltonian $H_\text{vq}$ in Eq.~\eqref{eq:electricfield}.
Substituting $E_x(t) = E_\text{ac} \sin \omega t$, 
the first term in the square bracket is rendered as
a transverse driving term 
$e E_\text{ac} \xrme \tau_1 \sin \omega t$.
Upon 
resonant driving $ \omega = \omega_\text{vq}$,
this term induces coherent Rabi oscillations of the qubit.
The speed of these
Rabi oscillations is characterized by the Rabi frequency 
\bean
f_\textrm{Rabi} = e E_\textrm{ac} \braket{\text{e}|x|\text{g}}/h.
\eean
The dependence of $f_\textrm{Rabi}$ 
on $x_0$,
for a moderate drive amplitude $E_\textrm{ac} = 1000$ V/m,
is shown as the solid red (gray) line in Fig.~\ref{fig:rabirelaxation};
the peak value above
$10^9$ Hz corresponds to sub-nanosecond single-qubit gates.

\subsection{Cavity-mediated $\sqrt{i \text{SWAP}}$ gate between two valley qubits}
\label{sec:cavity}

Electrically driven valley resonance is enabled by the
transverse coupling  between the valley qubit and the
electric field, that is, the first term in the square bracket
in Eq.~\eqref{eq:electricfield}.
The same term allows to realize a $\sqrt{i \text{SWAP}}$
logical gate on two valley qubits, if both are 
interacting with an empty mode of an electromagnetic cavity\cite{Zheng_twoqubitgate,Blais}.
Together with single-qubit operations, this two-qubit 
gate forms a universal gate set\cite{Barenco,Loss_PRA98}.
For the time of performing the logical gate, 
the two valley qubits has to be tuned on resonance with each other,
$\omega_\text{vq1} = \omega_\text{vq2} = \omega_\text{vq}$, 
and slightly detuned from the eigenfrequency of
the considered cavity mode $\omega_\text{cav}$.
The detuning should exceed the qubit-cavity coupling 
strength $e E_\text{cav} \xrme$,
where $E_\text{cav}$ is the cavity vacuum electric field
component along the $x$ axis. 
Assuming that the two valley qubits
are identical and feel the same cavity 
vacuum electric field,
the time required to perform the two-qubit gate is
\bean
\label{eq:sqrtswapgeneral}
t_{\sqrt{i \text{SWAP}}} = h \frac{\hbar |\omega_\text{vq}- \omega_\text{cav}|}{8 e^2 E^2_\text{cav} 
\xrme^2}.
\eean
For a numerical estimate of the gate time, we
take\cite{Tosi,Salfi_acceptorqubit} 
$E_\text{cav} = 30\ \text{V}/\text{m}$,
$\xrme = 12.4\ \text{nm}$, 
implying a qubit-cavity coupling strength 
of $e E_\text{cav} \xrme = 370 \ \text{neV}$,
that is equivalent to a rate of $\approx 90\ \text{MHz}$.
Then, by choosing the qubit-cavity 
detuning as 
$ |\omega_\text{vq}-\omega_\text{cav}| = 2\pi \cdot 720 \ \text{MHz}$,
from Eq.~\eqref{eq:sqrtswapgeneral} we find
$t_{\sqrt{i \text{SWAP}}} \approx 11 \ \text{ns}$.

\subsection{Valley-qubit relaxation via phonon emission}
\label{sec:relaxation}

Besides allowing for fast coherent control, 
the relaxation matrix element
$\braket{\text{e}|x|\text{g}}$, together with $\braket{\text{e}|z|\text{g}}$,
also exposes 
the valley qubit to relaxation processes induced by 
electrical potential fluctuations. 
Here, we focus on the example of
spontaneous phonon emission: 
the excited valley qubit can emit a phonon that carries 
away the qubit-splitting energy, 
and thereby the qubit relaxes to its ground state.
The relaxation process is characterized by 
the rate $\Gamma$.  
A practical question concerns the ratio of the achievable 
coherent Rabi frequency and the qubit relaxation rate: the former
should be much greater than the latter to have a functional qubit. 

The phonon-emission-mediated relaxation process between 
two electronic states in a  silicon QD
is described quantitatively in Ref.~\onlinecite{Tahan2014}:
Eq.~(6) therein is a formula for the relaxation rate $\Gamma$,
which is based on the dipole approximation. 
In our notation, and in the zero-temperature limit, 
that formula reads
\bean
\label{eq:gamma}
\Gamma = 
\frac{\omega_\text{vq}^5}{\hbar \pi \rho}
\left(
\xrme^2 \Upsilon_{xy} + \zrme^2 \Upsilon_z
\right),
\eean
where
\begin{subequations}
\bean
\Upsilon_{xy} &=& 
\frac{35 \Xi_d^2 + 14 \Xi_d \Xi_u + 3 \Xi_u^2}{210 v_l^7}
+
\frac{2\Xi_u^2}{105 v_t^7} \\
\Upsilon_z &=&
\frac{35 \Xi_d^2 + 42 \Xi_d \Xi_u + 15 \Xi_u^2}{210 v_l^7}
+
\frac{\Xi_u^2}{35 v_t^7}.
\eean
\end{subequations}
Here, the following notation is used 
for the material parameters of silicon: 
$\rho = 2330\ \text{kg}/\text{m}^3$ is the mass density,
$\Xi_d = 5 \ \text{eV}$ ($\Xi_u = 8.77 \ \text{eV}$) is the dilational (uniaxial) deformation potential, 
and 
$v_l = 9330 \ \text{m}/\text{s}$ 
($v_t = 5420 \ \text{m}/\text{s}$) 
is the longitudinal (transverse) sound velocity. 
Note that in this approximation, $\Gamma$ is proportional to 
the 5th power of the qubit splitting $\omega_\text{vq}$.  
Recall that in our case, 
the $y$-directional relaxation matrix element is zero, 
see Sec.~\ref{sec:interaction}.

Using our numerically computed 
relaxation matrix elements (Fig.~\ref{fig:matrixelements}a), 
we evaluate $\Gamma$ from
Eq.~\eqref{eq:gamma},
and show the result as the black 
dashed line in Fig.~\ref{fig:rabirelaxation}.
The key features are as follows. 
(i) If the step edge is far from the center of the QD ($|x_0| \gtrsim 3L$),
then $\Gamma$ is small, of the order of kHz, 
and it is independent of $x_0$. 
(ii) As the wave function overlaps more with the step 
 ($|x_0| \lesssim 3L$), 
 $\Gamma$ increases with orders of magnitude, 
 and grows above 100 MHz.
This is due to the 
large relaxation matrix element that peaks around 
the anticrossing point $x_0 \approx 0.5 L$ and arises
from the valley-to-charge conversion
and DQD-type behavior induced by the step.
(iii) Somewhat counterintuitively, $\Gamma(x_0)$ develops a small dip 
around the anticrossing point, 
where the relaxation matrix element has a peak.
An interpretation of this dip is obtained by
recalling the fact that $\Gamma$ is proportional to the 
5th power of the energy splitting of the qubit\cite{Tahan2014}, 
and the latter has a minimum at the anticrossing point
(see Fig.~\ref{fig:wfn}a).

We also compute the relaxation rate $\Gamma$ 
exactly, that is, without making the 
dipole approximation,
see Appendix \ref{app:relaxation} for details. 
The result is shown as the solid black line in Fig.~\ref{fig:rabirelaxation}. 
The exact $\Gamma$ is in general smaller than 
the dipole-approximated one. 
This is attributed to the phonon bottleneck effect\cite{Tahan2014}. 
Note that the best correspondence between the exact and
dipole approximated results is achieved in
the vicinity of the anticrossing
point $x_0 \approx 0.5 L$;
this is expected, as the qubit splitting is minimal here, hence the
wavelength of the emitted phonon is maximal, and therefore the 
ratio of the phonon wavelength and the lateral dot size,
characterizing the accuracy of the dipole approximation,
is maximal.

Finally, we note that Fig.~\ref{fig:rabirelaxation} 
suggests that it is possible to perform many single-qubit
operations within the relaxation time of the valley qubit,
if the step and the QD overlaps;
in particular, at the anticrossing point, 
$f_\text{Rabi}/\Gamma \approx 10^3$.

\subsection{Valley-qubit dephasing due to charge noise}
\label{sec:dephasing}

Besides the relaxation process due to electron-phonon interaction,
another mechanism of information loss for the valley
qubit is dephasing due to fluctuations of the external
electric fields.
For brevity, we refer to these fluctuations as charge noise. 
Charge noise can arise, e.g., as a consequence of fluctuating
gate voltages or charge traps in the nanostructure. 
Here, we discuss the relation between the strength
of charge noise and the inhomogeneous dephasing
time $T_2^*$ of the valley qubit. 

Aiming at order-of-magnitude estimates, we adopt a simple
model of charge noise: we assume that the corresponding
electric field
$\delta \vec E = (\delta E_x, \delta E_y, \delta E_z)$
is random, but 
homogeneous and quasistatic.
Dephasing arises, because 
the random electric field $\delta \vec E$ induces a
shift $\delta \omega_\text{vq} =
\omega_\text{vq}(\delta \vec E) - \omega_\text{vq}(\delta \vec E=0)$
in the valley-qubit energy splitting. 
The  $y$ component $\delta E_y$ of the random electric field
does not modify $\omega_\text{vq}$, because the step
is assumed to have translational invariance along $y$ and the
homogeneous  $\delta E_y$ does not change the shape of the 
parabolic confinement along $y$. 
The effects of the 
$x$ and  $z$ components are discussed separately below. 

\begin{figure}
\begin{center}
\includegraphics[width=1\columnwidth]{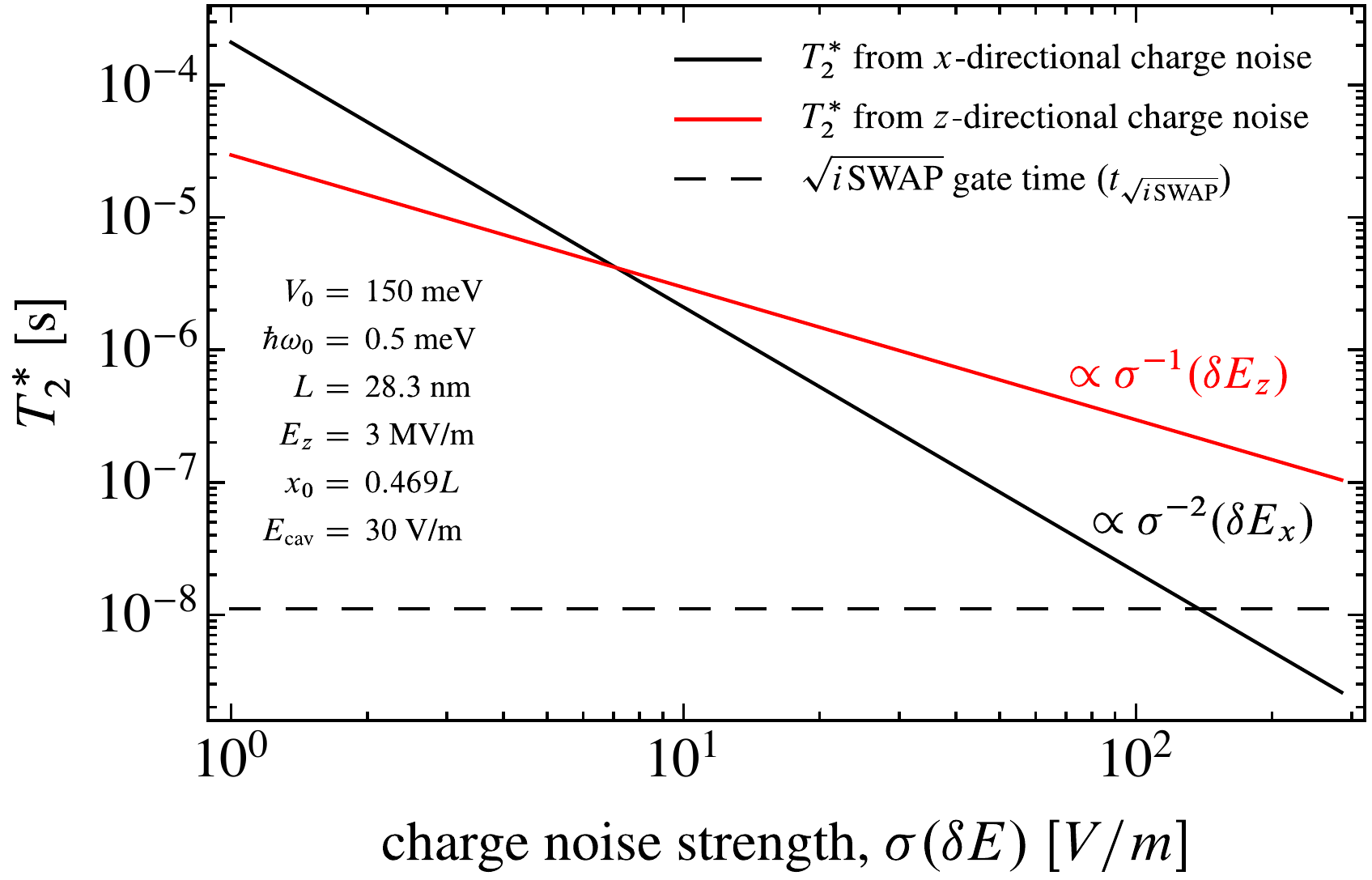}
\end{center}
\caption{(Color online) 
Valley-qubit dephasing time due to electric-field
fluctuations.
Black/red (gray) solid: inhomogeneous dephasing time
as a function of the strength (standard deviation) of the
$x$-directional/$z$-directional electric-field fluctations.  
For comparison, the black dashed line shows the 
cavity-mediated $\sqrt{i\text{SWAP}}$ gate time, 
for a vacuum cavity field $E_\text{cav} = 30\ \text{V/m}$.
\label{fig:swapdephasing}}
\end{figure}

The $x$ component $\delta E_x$ does induce a finite
$\delta \omega_\text{vq}$. 
In fact, the presence of $\delta E_x$ shifts the 
$x$-directional lateral confinement potential,
which 
is equivalent to 
shifting the step edge position,
which implies a change in the valley qubit splitting as
shown in Fig.~\ref{fig:wfn}a.
For weak noise, $\delta \omega_\text{vq}$ can be expressed
from the $x$-directional dephasing matrix element as
\bean
\delta\omega_\text{vq}=e\delta E_z\left(\xdme\right)/\hbar.
\eean
Note that the anticrossing point is a dephasing sweet spot 
with respect to $x$-directional charge noise, 
since the dephasing matrix element vanishes here. 
There, the relation between $\delta \omega_\text{vq}$ and 
the random electric field is expressed from a second-order
expansion as 
\bean
\delta \omega_\text{vq}(\delta E_x, 0, 0 ) = \alpha\, \delta E_x^2,
\eean
where
\bean
\alpha = \frac{1}{2}\frac{\partial^2 \omega_\text{vq}(\delta \vec E=0)}{\partial x_0^2}\left(\frac{eL^2}{\hbar \omega_0}\right)^2.
\eean
From the numerical data shown in Fig.~\ref{fig:wfn}a,
we obtain $\partial^2 \omega_\text{vq}(\delta \vec E = 0)/\partial x_0^2
\approx 3.72\times10^{27}\ \text{Hz}/\text{m}^2 $ at the anticrossing point, 
and, from that, we find 
$\alpha \approx 4.75\times10^3\ \text{Hz}/(\text{V}/\text{m})^2$. 
Then, we can identify the inhomogeneous dephasing time $T_2^*$
with the inverse of the typical noise-induced Larmor-frequency
detuning, $T_2^* \approx [\alpha \sigma^2(\delta E_x)]^{-1}$,
where $\sigma(\delta E_x)$ denotes standard deviation of $\delta E_x$.
This dependence is shown as the black solid line in 
Fig.~\ref{fig:swapdephasing}.

The  $z$ component $\delta E_z$ of the random electric field also induces 
a finite $\delta \omega_\text{vq}$.
For weak noise, 
is expressed as 
\bean
\delta\omega_\text{vq}=e\delta E_z ( \zdme )/\hbar,
\eean
and hence the dephasing time associated to these
$z$-directional charge noise is estimated as
$T_2^* = [\beta \sigma(\delta E_z)]^{-1}$, where
\bean
\beta = e \left| \zdme \right|/\hbar.
\eean
With the concrete parameter values corresponding to 
the anticrossing point of Fig.~\ref{fig:wfn}, we find
$\zdme = 2.22\times10^{-11}\ \text{m} $ and
$\beta = 3.37\times10^{4}\ \text{Hz}/(\text{V}/\text{m}) $.
The resulting relation between $T_2^*$ and $\sigma(\delta E_z)$ 
is shown as the red (gray) line in Fig.~\ref{fig:swapdephasing}.
 
In Fig.~\ref{fig:swapdephasing}, we compare how $T_2^*$
is influenced by the $x$-directional and $z$-directional 
components of charge noise.
For comparison, we also show, as the dashed horizontal 
line, the two-qubit gate time 
$t_{\sqrt{i\text{SWAP}}} \approx 11$ ns
estimated in section \ref{sec:cavity} (dashed horizontal line). 
These results suggest that in order to be able to perform 
at least a few ($\sim 10$) two-qubit operations within the 
inhomogeneous dephasing time, the charge noise strength
along $x$ ($z$) should be kept below
40 V/m (200 V/m).

\subsection{Relation to other models and real heterostructures}

Our results are based on a model\cite{Gamble2013}
where the atomic layers perpendicular to 
the heterostructure growth direction are represented by
continuous planes in which the electrons are described via
envelope functions, 
and a tight-binding description\cite{Friesen_PRB04} accounts for 
tunnelling between these atomic layers.
Within this framework, we provided 
a clear physical interpretation of our numerical results,
in section \ref{sec:valleytocharge}.
This interpretation is based on the condition that
for a step-free silicon/barrier interface, 
the charge densities of the ground and excited 
valley-qubit basis states are different
on the first atomic layer of silicon. 
If this condition is satisfied in other, 
potentially more realistic, 
models (e.g., accounting for multiple electronic bands, 
real-space structure of the electronic wave function,
disorder effects in the barrier material, etc)  and in real 
heterostructures, then we expect that 
the conclusions drawn from the model used here 
remain true at least on a qualitative level.

\section{Conclusions}\label{sec:sum}
We have analysed the influence of a single-atom high 
barrier step on the dynamics of a single electron confined to 
a silicon quantum dot.
We were focusing on the spectral and dynamical characteristics
of the single-electron valley qubit, that is, the 
two lowest-energy orbital states of the QD. 
We have found that placing the quantum dot over the
step has a strong influence on the properties of the valley
qubit.
(i) The wave functions of the two valley-qubit basis states are
deformed differently, leading to a mechanism of valley-to-charge
conversion, potentially useful for non-demolition readout of the
valley qubit. 
(ii) The presence of the step, together with an ac electrical 
excitation (induced by one of the confinement gates, for example),
can be utilized for resonant control of the valley qubit (electrically
driven valley resonance). 
(iii) Due to the step-induced enhancement of the 
interaction between the valley qubit and the external electric fields, 
two-qubit interactions can be mediated by an electromagnetic cavity. 
(iv) We have demonstrated that the valley-qubit relaxation rate
can be enhanced by  orders of magnitude  in the vicinity of the
interface step. 
(v) In conjunction with the valley-to-charge conversion
mechanism, we have demonstrated that a dephasing sweet spot 
against lateral ($x$-directional) electric field noise can be found if the 
relative location of the quantum dot and the step edge is 
set appropriately. 
Furthermore, we provided estimates for the inhomogeneous 
dephasing time caused by lateral ($x$-directional) and vertical 
($z$-directional) electric field fluctuations. 

These results provide insight to the fundamental dynamical
processes associated to the valley degree of freedom 
in imperfect silicon quantum dots, and an 
initial assessment of how the functionality 
of a valley qubit is influenced by the presence of
a barrier step at the silicon/barrier interface. 
Besides that, we think that the results presented here
will also contribute to the understanding of spin-qubit dynamics
in silicon quantum dots, which is often strongly influenced by
the valley degree of freedom.

\begin{acknowledgments}
We thank S.~Coppersmith, M.~Eriksson, M.~Friesen, A.~Dzurak, W.~Huang, M.~Veldhorst, N.~Zimmerman and R.~Joynt
for useful discussions.
We acknowledge funding from 
the EU Marie Curie Career Integration Grant CIG-293834,
OTKA Grants No.~PD 100373 and 108676, 
the Gordon Godfrey Bequest, 
and the EU ERC Starting Grant 258789.
A.~P.~was supported by the 
J\'anos Bolyai Scholarship of the Hungarian Academy of Sciences.
\end{acknowledgments}

\appendix

\section{Further details of the model}
\label{app:model}

In section \ref{sec:setupandmodel}, we specify the model
describing the energies and wave functions of the 
valley-qubit basis states $\ket{\text{g}}$ and $\ket{\text{e}}$. 
Here, we provide a few further details of the model and 
the numerical procedure. 

(1) 
The triangular quantum well along $z$, hosting the QD,
is modelled using a double-barrier structure.
The site index $j$ runs between -49 and 
134, the upper barrier
(shown in Fig.~\ref{fig:setup}) 
is the region $j \in \{-49,\dots,0\}$, 
the silicon quantum well is the region $j\in \{1,\dots,74\}$, 
and the lower barrier
(not shown in Fig.~\ref{fig:setup})
is the region $j \in \{75,\dots 134\}$.
Correspondingly, the function $\chi_b$, introduced in
section \ref{sec:setupandmodel} after Eq.~\eqref{eq:potentials},
representing the spatial range of the barrier material, 
is specified as 
\bean
\chi_b(x,y, -49 \leq j < 1) &=& 1, \\
\chi_b(x,y,j=1) &=& \Theta(x_0 - x),\\
\chi_b(x,y, j \leq 1 < 75) &=& 0, \\
\chi_b(x,y, 75  \leq j \leq 134) &=& 1.
\eean

(2)
To obtain the energy eigenvalues and wave functions in 
the presence of the interface step, we use the following 
procedure. 
First, we consider the case when the interface step is 
absent, and we 
numerically diagonalize the $z$-directional tight-binding
Hamiltonian $K_\text{chain} + V_i + V_b(x=0,y=0)$.
The obtained eigenvectors $\varphi_{n_z}$ ($n_z = 0,1, \dots, 173$), 
together with the harmonic-oscillator
eigenstates $\Psi_n$ ($n = 0,1, \dots$), 
provide a product basis 
$\psi_{n_x,n_y,n_z}(x,y,j) = \Psi_{n_x}(x) \Psi_{n_y}(y) \varphi_{n_z}(j)$,
which is the eigenbasis of the complete Hamiltonian $H$.
Then, in the presence of the interface step, 
the complete Hamiltonian $H$ 
is expanded in the truncated product basis, where $n_x \leq 14$ and
$n_y = 0$, and the resulting matrix is diagonalized
numerically. 
Note that it is sufficient to keep a single $y$-directional 
harmonic-oscillator eigenstate in the truncated basis, since
the interface step has translational invariance along $y$.

\section{Interpretation of the wave-function
patterns in Fig.~\ref{fig:wfn}b}
\label{sec:interpretation}

Here, we provide an interpetation of the wave-function patterns
(i) and (ii), discussed in section \ref{sec:valleytocharge}.

We start from the standard assumption 
of the envelope-function approximation\cite{Saraiva_EMA_PRB11} that $\ket{\text{g}}$ and
$\ket{\text{e}}$ are orthogonal linear combinations of two similar wave 
packets $\ket{\psi_{\pm k_0}}$
that are localized in momentum space
in the $z$ and $\bar{z}$ valleys, respectively:
\begin{subequations}
\label{eq:variational}
\bean
\label{eq:variationala}
\ket{\text{g}} &=& \frac{1}{\sqrt{2}} \left(
	e^{i\phi/2} \ket{\psi_{+k_0}} + e^{-i\phi/2} \ket{\psi_{-k_0}}
\right),
\\
\ket{\text{e}} &=& \frac{1}{\sqrt{2}} \left(
	e^{i\phi/2} \ket{\psi_{+k_0}} - e^{-i\phi/2} \ket{\psi_{-k_0}}
\right),
\label{eq:variationalb}
\eean
\end{subequations}
where 
\bean
\braket{j | \psi_{\pm k_0}} = F(z_j) e^{\pm i k_0 z_j}.
\eean
Here, $F(z)$ is the envelope function, which is spatially slowly 
varying, ensuring that $\ket{\psi_{\pm k_0}}$ are indeed
localized in the two valleys. 
The phase $\phi$ can be regarded as a variational parameter,
to be determined by the condition that the energy expectation
value of $\ket{\text{g}}$ should be minimal.

Importantly, the wave functions of Eq.~\eqref{eq:variational}
show sinusoidal spatial oscillations with wave number
$k_0$, as seen also in Fig.~\ref{fig:wfn}b.
Between neighboring lattice sites (distance $a$), 
the phase of that oscillation changes by 
$k_0 a = 0.82\, \pi /2$, a value close to $\pi/2$.
This explains why in Fig.~\ref{fig:wfn}b,
the quasi-node of $\psi_\text{g}$ at the last barrier layer 
is followed by a quasi-maximum at the first silicon layer
[see (i) in section \ref{sec:valleytocharge}].
This pattern of the wave function 
leads to a minimized potential-energy expectation value: 
having a wave-function 
quasi-node at the last barrier layer strongly reduces the 
potential-energy contribution of barrier, 
and having a wave-function quasi-maximum at the first silicon
layer, which is at the minimum of the $z$-directional confinement
potential, is also beneficial.

Finally, the relative phase of $\pi$ between 
the superpositions in Eq.~\eqref{eq:variationala}
and Eq.~\eqref{eq:variationalb} implies
that the spatial oscillations of $\ket{\text{e}}$
are phase-shifted with respect to those of $\ket{\text{g}}$ 
by $\pi/2$. 
Therefore, the wave function of $\ket{\text{e}}$ is peaked
at the last barrier layer but is close to zero 
at the first silicon layer [see (ii) in section \ref{sec:valleytocharge}].

\section{Valley relaxation}
\label{app:relaxation}

Here, we describe how we calculate the valley relaxation 
rate $\Gamma$, discussed in section \ref{sec:relaxation}, 
and shown in Fig.~\ref{fig:rabirelaxation} as the black solid (exact) and
dashed (dipole-approxated) lines.

We start from the zero-temperature Fermi's Golden Rule:
\bean
\label{eq:fgr}
\Gamma = 
\frac{2\pi}{\hbar} 
\sum_{\vec q,\lambda}
\left| 
\braket{\text{g}, \vec q \lambda 
| H_\text{eph} |\text{e}, 0}
\right|^2 
\delta(\hbar \omega_\text{vq} - \hbar v_\lambda q).
\eean
Here, bras and kets represent  joint states of the
composite electron-phonon system,  
$0$ denotes the vacuum of phonons,
and 
$\vec q$ ($\lambda$) is the wave number 
(polarization index) of the emitted phonon.
As for the electron-phonon interaction, we 
consider the deformation-potential mechanism, 
and describe it via the Herring-Vogt 
Hamiltonian:\cite{Herring,Yu_Cardona}
\bean
\label{eq:herringvogt}
H_\text{eph}
=
\Xi_d \text{Tr} (\varepsilon) + \Xi_u \varepsilon_{zz}.
\eean
Here, 
$\Xi_d$ is the dilational deformation potential,
$\Xi_u$ is the uniaxial deformation potential and
$\varepsilon$ is the strain tensor.
This form of $H_\text{eph}$ follows from 
the assumption that the the valley population of the
electronic wave function in the QD resides
in the $z$ and $\bar{z}$ valleys only.

The diagonal elements of the strain tensor,
that is, the elements that determine  $H_\text{eph}$
via Eq.~\eqref{eq:herringvogt},
read
\bean
\label{eq:strainunexpanded}
\varepsilon_{\alpha \alpha} =
i \sqrt{\frac{\hbar}{2 \rho V}} 
\sum_{\vec q,\lambda} 
\frac{e_{\vec q \lambda \alpha} q_\alpha}{\sqrt{ v_\lambda  q}}
e^{i \vec q \cdot \vec r}
\left(
	a_{\vec q,\lambda} + a^\dag_{-\vec q,\lambda}
\right).
\eean
Here, $\alpha \in \{x,y,z\}$,
$V$ is the sample volume and 
$\vec e_{\vec q \lambda}$ is the polarization vector 
of the phonon with wave number $\vec q$ 
and polarization index $\lambda \in \{l,t,t'\}$.

Note that from Eq.~\eqref{eq:strainunexpanded} it follows that 
transverse phonons do not contribute to the first term of the
electron-phonon Hamiltonian $H_\text{eph}$
in Eq.~\eqref{eq:herringvogt}.
Furthermore, we define the set of $t'$ phonons
such that their polarization vector lies in the $xy$ plane. 
That ensures that the $t'$ phonons do
not contribute to $H_\text{eph}$ at all. 

To obtain the valley relaxation rate $\Gamma$, 
we start from Fermi's Golden Rule \eqref{eq:fgr}, 
convert the sum for $\vec q$ to an integral
in spherical coordinates $(q, \theta_q, \phi_q)$, 
and perform the radial ($q$) integral.
This procedure yields
\begin{align}
\Gamma = 
\frac{\omega_\text{vq}^3}{8\pi^2 \hbar \rho}
\left(
\frac{
	\Xi_d^2 I_0
	+ 
	2 \Xi_d \Xi_u I_2
	+
	\Xi_u^2 I_4
	}{v_l^5}
+
\frac{\Xi_u^2 J}{v_t^5}
\right),
\end{align}
where
\begin{subequations}
\label{eq:IJ}
\begin{align}
I_n &=
\int_0^{2\pi} d\phi_q
\int_0^\pi d \theta_q
\sin \left( \theta_q \right)
\cos^n \left( \theta_q \right)
\left|
	\braket{g | e^{i \vec q_l \cdot \vec r} |\text{e}}
\right|^2,
\\
J &=
\int_0^{2\pi} d\phi_q
\int_0^\pi d \theta_q
\sin^3 \left( \theta_q \right) 
\cos^2 \left( \theta_q \right)
\left|
	\braket{g | e^{i \vec q_t \cdot \vec r} |\text{e}}
\right|^2,
\end{align}
\end{subequations}
where 
\bean
\vec q_\lambda = 
\frac{\omega_\text{vq}}{v_\lambda}
\threevector
	{\sin \left( \theta_q \right) \cos  \left( \phi_q \right)}
	{\sin \left( \theta_q \right) \sin  \left( \phi_q \right)}
	{\cos \left( \theta_q \right)}.
\eean
To obtain the exact valley relaxation rate, shown
in Fig.~\ref{fig:rabirelaxation} as the black solid line, 
we calculate these integrals numerically,
using the rectangle rule and a $15\times 15$
grid in the integration range 
$(\phi_q,\theta_q) \in [0,2\pi] \times [0,\pi]$.
To obtain the dipole-approximated result
\eqref{eq:gamma},
shown in Fig.~\ref{fig:rabirelaxation} as the black 
dashed line, 
the dipole approximation 
$e^{i \vec q_\lambda \cdot \vec r} \approx 1 + i \vec q_\lambda 
\cdot \vec r$ is used in Eq.~\eqref{eq:IJ}, 
allowing for an analytical evaluation of the angular 
integrals.

\bibliography{refs_Si}

\end{document}